# The Content of Statistics and Data Science Collaborations: the $Q_1Q_2Q_3$ Framework

Ilana M. Trumble[1], Jessica L. Alzen[2], Leanna L. House[3], and Eric A. Vance[1]

[1] Laboratory for Interdisciplinary Statistical Analysis, University of Colorado Boulder, Colorado, USA

[2] Center for Assessment, Design, Research and Evaluation, University of Colorado Boulder, Colorado, USA

[3] Department of Statistics, Virginia Tech, Blacksburg, Virginia, USA

## Abstract

For today's applied statisticians and data scientists, collaboration is a reality. Statisticians (and data scientists) may collaborate with domain experts across academic fields, industry sectors, and governmental and non-governmental organizations. Thus, statisticians must develop skills and techniques for collaboration. To this end, we advance a framework called the *Qualitative-Quantitative-Qualitative (*$Q_1Q_2Q_3$, pronounced "Q-Q-Q") approach to systematize the content of statistical collaborations. The $Q_1Q_2Q_3$ approach explicitly emphasizes the importance of the qualitative context of a project, as well as the qualitative interpretation of quantitative findings. We explain the $Q_1Q_2Q_3$ approach and each of its components as applied to statistics and data science consultations and collaborations. We provide guidance for implementing each stage of the approach and present data evaluating the effectiveness of teaching the $Q_1Q_2Q_3$ approach to beginning collaborators.

# 1. Introduction

Modern day statisticians and data scientists collaborate with experts from a broad range of fields, for example, across industry, academia, or government. According to Vance (2020), one of the end goals of a collaborative statistician is to make a deep contribution to the domain expert's field. To achieve this goal, statisticians must develop skills and techniques for collaboration. Bhattacharyya (2017) states, "Applied statisticians and data scientists must be - almost by definition - collaborative because they rarely author the studies they design or generate the data they analyze…to meaningfully collaborate requires multi-facetted skills and a lot of practice" (p. 6). Statisticians armed with collaboration skills have the potential to create new knowledge they could not create otherwise (Love et al., 2017).

Vance and Smith (2019) introduced the ASCCR framework, which outlines five essential components of a successful collaboration: Attitude, Structure, Content, Communication, and Relationship. In this paper, we develop and detail practical approaches to the ASCCR framework's Content of collaborations. Specifically, we advance a framework called the *Qualitative-Quantitative-Qualitative (*$Q_1Q_2Q_3$, pronounced "Q-Q-Q") approach to systematize the content of statistical collaborations. As the name conveys, the $Q_1Q_2Q_3$ approach includes both qualitative and quantitative components. Ograjenšek and Gal (2016) write that all research is driven by the qualitative 'need to know' that exists in real-world situations or in scientific inquiries. Ograjenšek and Gal (2016) state, "The 'need to know' dictates how a certain problem may be dealt with: via quantitative approach and by application of statistical methods (only), via the use of a qualitative approach and related methods (only) or through mixed-methods research, that is, by a combination of quantitative and qualitative reasoning and methods" (p. 174). The $Q_1Q_2Q_3$ framework presents the qualitative 'need to know' as key to the Content of any statistics or data science research collaboration.

The $Q_1Q_2Q_3$ approach is based on the eight elements of thought for critical thinking introduced by Elder and Paul (2013). Leman et al. (2015) expanded upon this framework and presented their own $Q_1Q_2Q_3$ approach for learning and teaching statistics and data analytics. Leman et al. (2015) state that in a data analysis project, analyzing the Qualitative issues ($Q_1$) of the context-specific question must precede any Quantitative ($Q_2$) mathematics or computations. Any quantitative findings must be Qualitatively ($Q_3$) summarized and assessed in a manner consistent with the questions asked in the $Q_1$ phase of the analysis. Leman et al. (2015)

conclude that the $Q_1$-$Q_2$-$Q_3$ format can be applied to individual lessons, courses, and entire degree programs.

While Leman et al. (2015) developed their $Q_1Q_2Q_3$ approach for learning and teaching statistics and data analytics, we present a $Q_1Q_2Q_3$ approach for interdisciplinary collaborations. Previously, Vance and Smith (2019) adopted the $Q_1Q_2Q_3$ approach as the primary method for training collaborative statisticians and data scientists in the Content of applied projects within the ASCCR framework. Vance and Smith (2019) found the $Q_1Q_2Q_3$ to be accessible, easy to teach, and beneficial for collaborations. In this paper, we further develop the $Q_1Q_2Q_3$ approach as an integral part of a theory for applied statistics.

There is potential for negative consequences if statistical collaborators do not complete each phase of $Q_1Q_2Q_3$. In general, statisticians are prone to rushing or skipping the qualitative stages, most likely because most formal statistics and data science training focus on quantitative skills. A statistician that does not give the qualitative stages their due diligence is at risk of making a Type III error, the error committed by giving the right answer to the wrong problem (Kimball, 1957). For example, say that a statistician learns that the domain expert is interested in comparing the size of ablations created by bipolar and monopolar radiofrequency devices used to destroy hepatic tumors (Dodd et al., 2017). The statistician is excited to see that the domain expert has taken many measurements, and that previous literature suggests ablation sizes follow a normal distribution. The statistician completes a two-sample *t*-test and is satisfied that the domain expert's question has been answered. However, the statistician later learns that it was actually the difference in the variance of the ablations that has clinical implications, not the difference in the average ablation size. While the statistician gave the correct answer to the question they thought was important, failure to diligently investigate the qualitative background led to the Type III error of producing the right answer to the wrong question.

The remainder of this paper proceeds as follows. In Section 2, we explain the $Q_1Q_2Q_3$ approach and each of its components as applied to statistical consultations and collaborations. In Section 3, we provide guidance for implementing $Q_1$, $Q_2$, and $Q_3$ in collaborative statistics and data science projects. In Section 4, we present data evaluating the effectiveness of the $Q_1Q_2Q_3$ approach for beginning collaborators. In Section 5, we discuss the implications of the $Q_1Q_2Q_3$ approach for the practice of statistics and directions for future work. In Section 6, we summarize with our conclusions.

## 2. Explanation of $Q_1Q_2Q_3$

The first stage, $Q_1$, stands for "Qualitative." In this stage, the statistician should think about the domain problem and make judgments as to the relevance of the data in hand before a model is formulated or data is analyzed. $Q_1$ includes the context of the problem, the importance of the problem to the field, and how the eventual solution will be used. $Q_1$ also includes considering the perspectives and backgrounds of both the domain expert and the statistician. In his 1997 American Statistical Association Fisher Memorial Lecture, Mallows' stated, "The main challenge of applied statistical work is that of taking proper account of contextual issues. Good techniques are not enough; nor are good computer programs, nor powerful theorems. A major intellectual attraction of the discipline is the subtlety of the interplay between the formal statistical procedures and the imperfectly understood substantive questions" (Mallows, 1998, p.3). Mallows (1998) defined the *"Zeroth Problem"* as "considering the relevance of the observed data, and other data that might be observed, to the substantive problem" (p. 2). $Q_1$ is where the statistician solves the *"Zeroth Problem"* and lays the qualitative foundation necessary to contribute to the domain expert's field.

The second stage, $Q_2$, stands for "Quantitative." In this stage, the statistician applies quantitative techniques to answer the domain problem. During $Q_2$, analysts design studies, explore and summarize data, formulate models, and perform statistical inference. Part of the $Q_2$ stage is deciding analytically whether the data collected are relevant for answering the questions from $Q_1$. Therefore, it is imperative that $Q_1$ is completed before moving on to $Q_2$.

The third and final stage, $Q_3$, stands for "Qualitative." In this stage, the statistician uses the results from $Q_2$ to answer the qualitative questions from $Q_1$. The statistician should interpret the quantitative findings in language accessible to the domain expert, and provide conclusions and recommendations for the solution or decision. On the results of applied statistics projects, Gal and Ograjenšek (2016) write: "Conclusions have to be presented or reported to clients or stakeholders in ways that they understand and find easy to make sense of, and be congruent with their 'policy language'" (p. 204). Additionally, the statistician should provide recommendations for actions based on evidence from $Q_2$. These recommendations should stem from an "evidence into action for development" mindset (Olubusoye et al, 2021). $Q_3$ also includes assessing the impact of any assumptions made in $Q_1$ or $Q_2$ on the qualitative generalizability of the conclusions. For example, if a study assumes that the population of interest is senior citizens from a specific retirement home and collects data accordingly, any qualitative conclusions made in $Q_3$ will only be applicable to that population of senior citizens. Lastly, as part of $Q_3$, statisticians and data scientists should reflect upon whether $Q_1$ and $Q_2$ were conducted ethically and responsibly.

Formally, we define $Q_1Q_2Q_3$ as the three qualitative and quantitative stages of critical thinking necessary for effective statistical collaborations. While the three Qs are defined in sequence, there is often iteration between them. For example, while exploring data in $Q_2$, it might make sense to update research questions in $Q_1$. Figure 1 depicts the stages of $Q_1Q_2Q_3$ and their potential relationships.

**Figure 1.** The $Q_1Q_2Q_3$ stages of critical thinking for effective statistical collaborations.

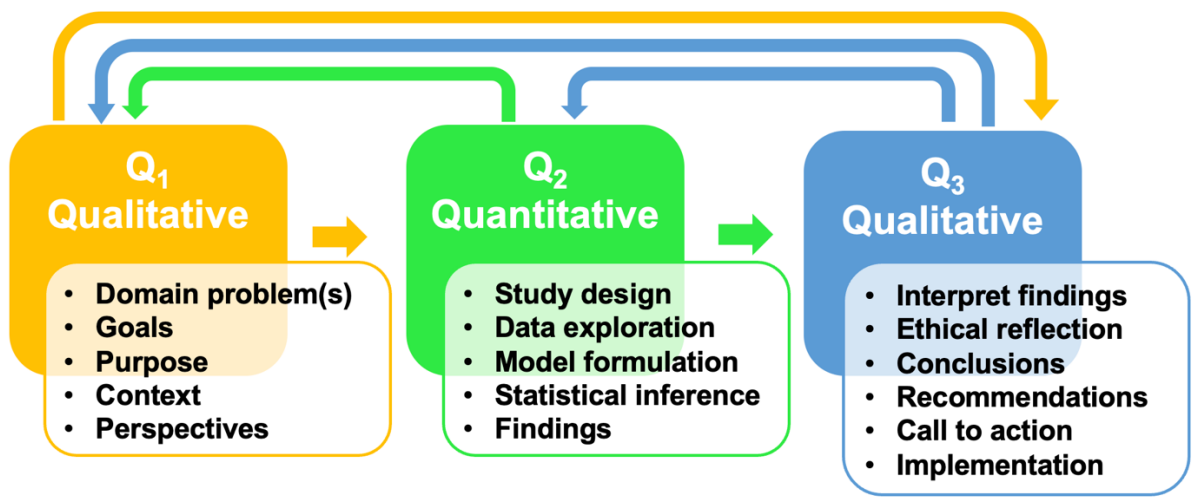

An analogy for the $Q_1Q_2Q_3$ process is the hero's journey (Campbell, 2003). In this analogy, a statistical collaboration is an epic adventure, and both the statistician and domain expert are the heroes. The hero's journey consists of three stages: Departure, Initiation, and Return. The three stages of the hero's journey parallel the three stages of $Q_1Q_2Q_3$, as follows.

1. *Departure.* In the Departure stage, which parallels $Q_1$, the heroes live in the ordinary world and receive a call to adventure. The ordinary world is the domain expert's field. The call to adventure is the problem that the domain expert presents to the statistician.
2. *Initiation.* In the Initiation stage, which parallels $Q_2$, the heroes traverse the threshold to an unknown world. There, the heroes face trials. In our analogy, the special world is the world of quantitative reasoning, and the trials are the quantitative analyses that must be performed.

3. *Return.* In the Return stage, which parallels $Q_3$, the heroes again traverse the threshold between the worlds, returning to the ordinary world with gained treasure which may be used for the benefit of others. The treasure is the knowledge that the statistician and domain expert have gained by embarking on the quest. The knowledge is used to create a call to action, which helps others and benefits society.

Through the journey, the heroes are transformed and gain wisdom over both worlds. Likewise, through the collaboration, the statistician and domain expert gain expertise in both the research domain and quantitative methods.

The $Q_1Q_2Q_3$ stages can be found in many influential research papers. For example, a team of statisticians and a historian collaborated to predict where enslaved people came from within Africa before they were forcibly shipped across the Atlantic (Wiens et al., 2022). Components of $Q_1Q_2Q_3$ from this study are:

- $Q_1$: Wars in the Kingdom of Oyo (modern-day Nigeria) from 1817–1836 resulted in 121,000 enslaved people sent in slave ships to the Americas. While historians have a good record of where the enslaved went across the Atlantic, they have no records of where they were from *within* Africa. The team compiled data on trade routes and the time, place, and severity of conflicts in Oyo.
- $Q_2$: The team used a decision process to simulate the transport of the enslaved to ports of departure. The researchers aggregated the simulations to predict the conditional probabilities of the likely origin locations of the enslaved. The origins of the enslaved were modeled and visualized (see bit.ly/Oyoorigins).
- $Q_3$: The team recommends that statisticians refine such models and historians collect more and better data. This research helps to better understand the history of Africa and the entire Atlantic world, whereby the ocean *connects*, rather than disconnects, Africa, the Americas, and Europe.

# 3. Implementing $Q_1Q_2Q_3$ in Statistical Collaborations

## 3.1 Implementing $Q_1$

To successfully implement $Q_1$, the statistician and data scientist should create shared understanding (Vance et al., 2022a) about the following six qualitative aspects of the research project. The answers to these questions should be conveyed and verified between the statistician/data scientist and the domain expert.

1. What is the domain problem?
2. Why is the problem important or interesting? And, to whom?
3. How will the eventual solution be used?
4. What potential data could solve the domain problem? (I.e., what data, if it were available and accessible, would help answer the underlying questions?)
5. The Five Ws and one H of the actual data, if any has been collected.
    a. What data was collected?
    b. Who or what collected the data?
    c. Why, and for what purpose, was the data originally collected?
    d. When was the data collected?
    e. Where was the data collected?
    f. How was the data collected?
6. What may be the qualitative relationships between variables, for variables both observed and unobserved?

To obtain this information and complete the $Q_1$ stage, Vance et al. (2022b) outline three useful strategies for asking great questions: 1) preface questions with their intent, 2) listen, paraphrase, and summarize, and 3) model and cultivate a collaborative relationship. For example, to discuss the importance of the domain problem (Question 2), the collaborator may preface their question and say, "Understanding your motivations and your reasons for researching this area helps me get excited about the research and really helps my brain think better statistically. So I'm curious, why do you want to answer this research question?" After the domain expert responds, the statistician could employ the listen, paraphrase, and summarize strategy by saying, "That's so interesting. Just to make sure I understand, this research is important because…"

Prefacing questions with their intent can address common hurdles the collaborator may face in the $Q_1$ stage. For example, sometimes, the domain expert may resist thoroughly discussing the information relevant to $Q_1$ because they may feel it is an inefficient use of time or unnecessary for the statistician to complete their task. Prefacing questions with their intent can address this hesitancy. For example, to start a discussion about the experimental design (Question 5, part f.), the statistician may say, "So I can better understand the experiment and model the data well, how were the treatments assigned?" By prefacing questions with the intent, the statistician explains to the domain expert why it is worth their time to discuss the $Q_1$ components of the problem.

Another strategy to encourage discussion about $Q_1$ aspects of the project is to include the $Q_1$ questions in the meeting agenda, especially when it is the first meeting between the statistician and domain expert. When the statistician begins the meeting with $Q_1$ questions already written down, there is structure and incentive for the domain expert to answer them. We provide an example shared meeting notes document, including an agenda that contains $Q_1$ questions, at [bit.ly/gdoccollabtemplate](bit.ly/gdoccollabtemplate). We encourage statistical collaborators to make use of this or a similar shared document in their own meetings.

Question 6 is essential for identifying confounders, mediators, and effect modifiers (a.k.a. moderators or interactions). A helpful strategy is to work with the domain expert to draw a causal diagram with arrows depicting the relationships between variables (Pearl, 1995). The three diagrams in Figure 2 depict simple examples of causal diagrams for a confounder, effect modifier, and mediator.

**Figure 2.** Example causal diagrams showing a) a confounder, b) an effect modifier, and c) a mediator. In all diagrams, the dependent variable/outcome is in the bottom right, the primary predictor of interest is in the bottom left, and the confounder, effect modifier, or mediator is on top. These diagrams are helpful tools for completing the $Q_1$ stage of collaboration.

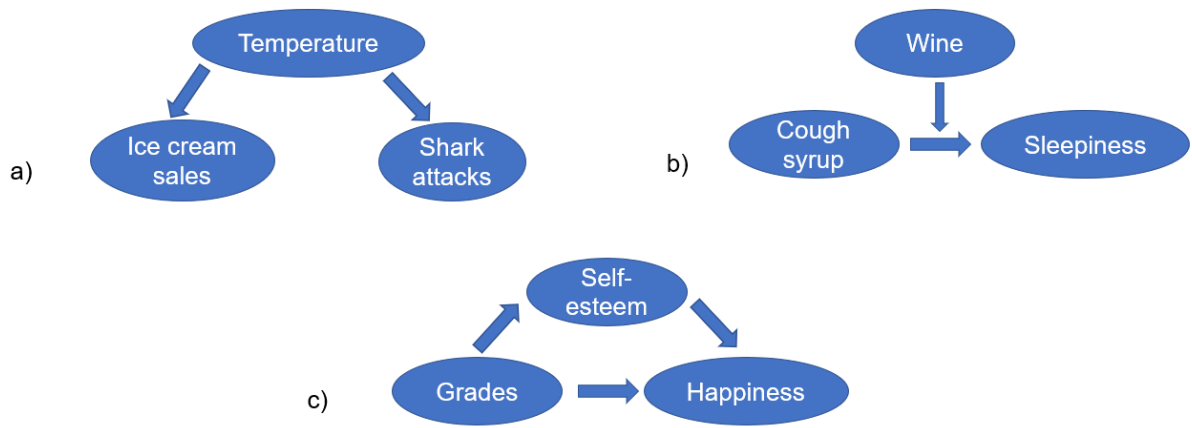

## 3.2 Implementing $Q_2$

Statisticians are likely most familiar with the $Q_2$ stage of $Q_1Q_2Q_3$. This stage is often the sole focus of standard statistics and data science courses. For this reason, we will not detail $Q_2$ methods in this paper. However, even though $Q_2$ is a technical step, nuanced choices are needed to decide what methods to apply and how to interpret/evaluate their success. We recommend quantitative analyses guided by three core principles of data science: Predictability, Computability, and Stability (Yu and Kumbier, 2020).

The Predictability, Computability, and Stability (PCS) principles for conducting data science were proposed by Yu and Kumbier (2020) to create a common culture and assure truthful findings from data. We briefly summarize each principle:
- Predictability: a model's ability to accurately predict new observations. Measures of model *prediction* can serve as tools for evaluating, improving, and applying analytic methods. Predictability serves as a reality check for the quantitative methods.
- Computability: a model or algorithm's computational efficiency and scalability. For example, available computing power may directly influence how and where data are collected, stored, shared, processed, and summarized. Computability ensures results are obtainable.

- Stability: whether another researcher making alternative, appropriate decisions would obtain similar conclusions. Stability ensures the reproducibility of results relative to human decisions.

The PCS principles describe how to implement $Q_2$ by providing a common value system among statisticians. Statisticians should consider, measure, and report each element of PCS in their $Q_2$ analysis. For specific suggestions for quantitative methods that may be used to measure each element, see Yu and Kumbier (2020). When all elements of PCS are considered, measured, and reported explicitly, analytic results have the potential to be "responsible, reliable, reproducible, and transparent … across fields of science" (Yu and Kumbier, 2020, p. 3920). While we discuss PCS in the context of $Q_2$, these principles should be upheld throughout the $Q_1Q_2Q_3$ process.

We note that adhering to PCS principles may look different for researchers with different backgrounds. For example, social scientists with data from human subjects may have lower expectations for model predictiveness than pharmacological researchers of clinical drug trials, yet both require clear standards for reproducible science. Similarly, data management and model applications might be computationally feasible for researchers with one set of resources, but infeasible for researchers with a different set of resources. Lastly, teams must consider whether the quantitative methods will produce results that are interpretable and meaningful to all team members. Collaborative efforts must negotiate all differences in PCS and articulate the common choices made to complete $Q_2$.

Notably, the $Q_1Q_2Q_3$ approach honors the principle of Stability for statistical collaborations by explicitly identifying the subjectivity of the statistician when conducting analyses. Successful statistical collaborations depend on careful critical thinking when making choices using personal perspectives and experience throughout the analytic process, rather than just accurate executions of computational algorithms and statistical models. The $Q_1Q_2Q_3$ formalizes this critical thinking by including humans explicitly in the process for learning from data. When Stability is upheld, a different statistical collaborator going through the $Q_1Q_2Q_3$ critical thinking process would reach the same $Q_3$ conclusions.

## 3.3 Implementing $Q_3$

The final stage of $Q_1Q_2Q_3$, $Q_3$, is crucial to every statistics collaboration. In $Q_3$, the statistician uses quantitative evidence from $Q_2$ to answer the qualitative questions from $Q_1$. Without $Q_3$, statistics and data science results will have no impact and will have been done in vain. As part of $Q_3$, statisticians should discuss the measures of PCS in language accessible to the domain expert. Lastly, the statistician should consider any ethical implications of the qualitative answers.

We outline seven questions for statisticians and data scientists to ask themselves as a guide to completing $Q_3$. The answers to these questions should be conveyed and verified between the statistician and the domain expert.

1. Qualitatively, what do the results mean?
2. What are the answers to the domain experts' questions?
3. Does the domain expert understand the relevance of the statistical answers to the research goals?
4. What are the implications of the answers, including ethical implications?
5. How can we visually display and communicate the results of the analysis in a way the domain expert and their stakeholders (i.e., advisor, boss, peers, etc.) will understand?
6. What are the constraints, limitations, and assumptions of the quantitative methods? What conditions are necessary for the results to be valid?
7. What are the actions that should be taken as a consequence of the results?

In the $Q_3$ stage, it is crucial that the statistician and domain expert have clear communication and develop shared understanding about the results. We outline five strategies for achieving shared understanding. First, results should be explained in language that is accessible to the domain expert. Second, the statistician should be intentional about providing the domain expert multiple opportunities to speak and ask questions throughout meetings. For example, periodically checking in with the domain expert by asking "What can I clarify?" or "What can I explain further?" often and after each explanation can be helpful. Third, the statistician can ask the domain expert how they would explain the results in their own words, so long as the intent is prefaced. For example, the statistician might say, "I want to make sure that I have explained these results clearly. To check that I have done that, could you tell me how you would explain these results?" Fourth, the statistician and domain expert can brainstorm together what actions or implications the results suggest. Lastly, all results, findings, and consequent actions or implications should be written up clearly in a document and shared between the domain expert and statistician.

Finally, throughout $Q_3$, the statistician will benefit from adopting the attitude of a collaborative relationship (Alzen et al., 2022). In doing so, the statistician considers themselves just one of many experts in the room. He or she believes that $Q_3$ is not about showing off their statistical expertise, rather, it is about helping the domain expert make a good decision. The statistician and domain expert are on the same team, and the statistician succeeds when the domain expert succeeds.

# 4. Assessing the Success of the $Q_1Q_2Q_3$ Approach

We use both quantitative and qualitative methods to provide evidence of the success of the $Q_1Q_2Q_3$ approach in the Statistical Collaboration course at the University of Colorado Boulder. Data include Likert-scale survey items and open-response reflection assignments. Our sample includes two groups of students: 33 students who took the course in Fall 2021 or Spring 2022, and 76 students who took the course between Fall 2016 and Spring 2021. Fall 2021 and Spring 2022 students completed a pre-post survey where responses were measured in the first and last week of class. Fall 2016 to Spring 2021 students completed only a post-survey, where responses were measured in May 2021 for all students. There was a 100% response rate on the Fall 2021 and Spring 2022 pre-post survey, and a 27.6% response rate (23 out of 76) on the Fall 2016 to Spring 2021 post-survey.

The same four prompts were included in both the pre- and post-surveys (Tables 1 and 2). These prompts reflect students' self-efficacy in skills necessary for the $Q_1Q_2Q_3$ approach. The first prompt describes $Q_1$ skills, the second prompt is part of $Q_2$, and the third and fourth prompts are elements of $Q_3$. Students rated the extent to which they agreed with the prompts on a Likert scale ranging from 1 (Strongly disagree) to 6 (Strongly agree). Table 1 shows the results from the post-survey for students from Fall 2016 to Spring 2021, and Table 2 shows the results from the pre-post survey for students from Fall 2021 and Spring 2022.

We see that the Fall 2016 to Spring 2021 students rated the prompts highly, on average, indicating that they either agreed or strongly agreed that they can effectively navigate the $Q_1Q_2Q_3$ approach in statistical collaborations. For the pre-post survey conducted for students enrolled in Fall 2021 or Spring 2022 (Table 2), higher scores were observed at the end of the course for all four prompts. Responses from the first and final week of class were compared using the paired Wilcoxon signed-rank test. Two prompts had a significant difference at the 0.05 significance level: "I can listen, paraphrase, and summarize to effectively achieve mutual understanding with another person" ($p$ = 0.035) and "I can structure and organize my work on projects (project management or statistical workflow) effectively" ($p$ = 0.0008). The other two prompts were not significant at the 0.05 level: "I can effectively interpret statistical results in disciplines other than my own" ($p$ = 0.24) and "I can effectively explain statistics and data science to people in disciplines other than my own" ($p$ = 0.071). The data are promising for future research with a larger sample of students, work that we intend to complete over the next two years.

**Table 1.** Fall 2016 through Spring 2021 student responses (N=23) to a post-course survey. Students rated the extent to which they agreed with the prompts on a 6-point Likert scale.

| Items scored from 1 (Strongly disagree) - 6 (Strongly agree) | Mean (SD) |
| --- | --- |
| I can listen, paraphrase, and summarize to effectively achieve mutual understanding with another person. | 5.52 (0.51) |
| I can structure and organize my work on projects (project management or statistical workflow) effectively. | 5.30 (0.63) |
| I can effectively interpret statistical results in disciplines other than my own. | 5.13 (1.01) |
| I can effectively explain statistics and data science to people in disciplines other than my own. | 5.26 (0.81) |

**Table 2.** Fall 2021 and Spring 2022 student responses (N=33) to a pre-post survey where responses were measured in the first and last week of class. Students rated the extent to which they agreed with the prompts on a Likert scale ranging from 1 (Strongly disagree) to 6 (Strongly agree).

|  | Pre-Mean (SD) | Post-Mean (SD) | Mean Difference (Post - Pre) | $p$-value* |
|---|---|---|---|---|
| I can listen, paraphrase, and summarize to effectively achieve mutual understanding with another person. | 5.03 (0.77) | 5.33 (0.69) | 0.30 | 0.035 |
| I can structure and organize my work on projects (project management or statistical workflow) effectively. | 4.70 (0.88) | 5.27 (0.63) | 0.58 | 0.0008 |
| I can effectively interpret statistical results in disciplines other than my own. | 4.45 (1.00) | 4.67 (0.89) | 0.21 | 0.24 |
| I can effectively explain statistics and data science to people in disciplines other than my own. | 4.42 (0.94) | 4.76 (1.00) | 0.33 | 0.071 |

* Computed using the Wilcoxon signed-rank paired test.

To further expand our body of evidence regarding the effectiveness of teaching the $Q_1Q_2Q_3$ approach in Statistical Collaboration, we asked the Fall 2021 and Spring 2022 students how valuable they expect the $Q_1Q_2Q_3$ approach to be throughout their careers. On a scale of 1 (not valuable at all) to 6 (extremely valuable), the mean (SD) student response was 4.70 (0.98), i.e., closer to "very valuable" than "moderately valuable." These findings imply that students generally expect the $Q_1Q_2Q_3$ approach to statistical collaborations will be useful to them in the future.

To gain a better understanding of why students feel this way, we turn our attention to qualitative data. Throughout the course of the semester, students submitted three individual reflections. In addition, students completed an end-of-semester reflection. We provide a selection of the responses in which students explicitly reflected on the $Q_1Q_2Q_3$ approach (Table 3). Five different students are represented in the five quotations. We see in these quotations that students consistently identify the $Q_1Q_2Q_3$ approach as useful in statistical collaboration projects and that they effectively learned how to use the strategy through the course. These responses are particularly notable because students were not directly prompted to reflect on the $Q_1Q_2Q_3$ approach in these assignments, but rather on everything they learned throughout the entire course.

**Table 3.** Fall 2021 quotes about effectiveness of $Q_1Q_2Q_3$ approach. Students were prompted to reflect on the entire course at the end of the semester. Students were not directly prompted to reflect on the $Q_1Q_2Q_3$ approach, yet they mentioned it in their reflections.

| Student quotations from end-of semester reflections |
| --- |
| [This project] is a classic case where the domain expert comes in with a method in mind, and we have to rewind to get to the substance – [The Domain Expert] started the project telling us she needed to do a "cluster analysis" (unsupervised learning), but as we dug deeper, we figured out that one of her committee members suggested that as a possible technique, when in reality other methods may be more appropriate. By using $Q_1Q_2Q_3$, we were able to figure out the actual goals of the project, and decided that an exploratory and descriptive analysis would be more appropriate. |
| The collaboration also further taught me how important it is to understand the $Q_1$ of a project. Doing so for this project not only helped us as collaborators to solve [the Domain Expert's] problem, but also the $Q_1$ questions we asked seemed to make the DE feel more sure about what exactly he was doing. |
| I personally believe the Content element is the most important to effective collaborations. The $Q_1Q_2Q_3$ setup ensures that both the statistical collaborator and the domain expert are talking about the same problem, and helps eliminate any misunderstandings about what the other is working on. |
| Another very useful skill I learned in this class was the focus on qualitative aspects of the research. In the past I have often jumped right into the statistical questions of a project, but my experience in this class has shown me the importance of the $Q_1Q_2Q_3$ framework. |
| LISA helped me keep a good set of meeting notes, and split the conversation into $Q_1$, $Q_2$, and $Q_3$ (which I think I would sometimes unknowingly do, but now that I know about it, I do it intentionally and better than before). |

We collected one additional piece of qualitative data from Spring 2022 students. In the final week of the course, we asked students to respond to the prompt, "Has learning about the $Q_1Q_2Q_3$ approach strengthened your ability to successfully complete statistics or data science collaboration projects? Why or why not?" (Table 4). Overall, we see that 10 out of the 11 students who responded thought that the $Q_1Q_2Q_3$ approach increased their ability to successfully collaborate on statistics projects. One student did not think it was useful but seemed to have a misunderstanding of what the stages of the $Q_1Q_2Q_3$ approach are.

**Table 4.** Spring 2022 student responses at the end of the semester of the Statistical Collaboration course.

| **Spring 2022 student responses to:** <br> **"Has learning about the Q1Q2Q3 approach strengthened your ability to successfully complete statistics or data science collaboration projects? Why or why not?"** |
|---|
| Yes, I think this approach has strengthened my ability to complete collaboration projects because it has forced me to slow down and fully understand the research goals before trying to interpret data. |
| I think it has because it gives a good framework to go about the work. It is beneficial to really flesh out the qualitative goals initially to make sure that subsequent quantitative work is worth the time (i.e. even answers the question). Having a thorough understanding of what the results mean is also beneficial. |
| I believe so. It has added additional structure to the problem solving process, which should serve to better complete statistics or data science collaboration projects in an effective manner. |
| I think it is a good structure for projects. I have not done a lot of projects without that structure so it is hard to compare. |
| Yes. In those projects, I need to collaborate with people who may not familiar with statistics/math and QQQ approach is very helpful in this situation |
| It is successful because it gives me to make the steps |
| I think it brings a much needed structure to my ad hoc style. This way I don't get lost in the weeds. |
| Yes. Using this approach helps me better understand the projects and helps me organize the work I will be doing. |
| I think it is helpful for figuring out what exactly you want to know before diving into existing data/ data collection. |
| Yes, it has reminded me to scale back to focus on the big picture at the end of meetings |

> Not really - I think for my purposes I would change the order to Q2, Q1, Q3. I think it's more natural to discuss the bigger qualitative picture before diving into the quantitative components of the project. If I ever work in industry and present to the C-suite, then I would consider Q1, Q2, Q3.

## 5. Discussion

This paper contributes to the growing research on the development of collaboration skills in statisticians and data scientists. Unlike previous research on statistical collaborations, this paper outlines specific elements for the content of statistical collaborations. We provide evidence of the effectiveness of the Qualitative-Quantitative-Qualitative process in statistical collaborations by combining both quantitative survey results with qualitative student reflections. These results provide well-rounded evidence for students' experiences with the $Q_1Q_2Q_3$ method, as well as adds to the body evidence surrounding the need to train students in statistical collaboration skills. The $Q_1Q_2Q_3$ strategies outlined herein are essential techniques for successfully understanding the domain problem, performing sound and reproducible quantitative analysis, and transforming the results into action for the benefit of society.

The study presented herein has some limitations. First, our data does not involve an objective evaluation of students' $Q_1Q_2Q_3$ skills, rather we rely solely on self-reflection. Future work will involve objectively assessing students' $Q_1Q_2Q_3$ skills from submitted project reports and videos of meetings with domain experts. Second, our data lack a control group, therefore we do not have sufficient evidence to support a strong causal claim for improvements seen over the semester. It is possible that other factors such as maturation or other student experiences contributed to the development of $Q_1Q_2Q_3$ skills. Future work will involve including a control group in the surveys. Third, for the students who completed only the post-survey, there is no comparison between their $Q_1Q_2Q_3$ skills at the start and end of the semester. Therefore, we cannot conclude that the course strengthened their skills. Lastly, for both the pre-post and post-survey, the sample sizes were limited (N=33 and N=23, respectively). Future work will involve collecting more pre-post data to increase the sample size. Future work will also involve a detailed qualitative analysis of additional student responses to expand the field in understanding the use of the $Q_1Q_2Q_3$ approach and teaching statistical collaboration broadly.

We believe that statistics and data science educators should consider the $Q_1Q_2Q_3$ approach as an integral part of teaching collaboration skills. The $Q_1Q_2Q_3$ approach provides a clear path for implementing the critical thinking stages necessary for successful collaborations. Not only will teaching collaboration skills with the $Q_1Q_2Q_3$ approach advance the field in training statisticians, but also will reduce the chances of committing a Type III error and strengthen the impact of statistics and data science in general. Professional statisticians can also benefit from implementing the $Q_1Q_2Q_3$ approach and practicing the $Q_1Q_2Q_3$ strategies in their collaborations.

# 6. Conclusion

This paper describes the $Q_1Q_2Q_3$ approach as a framework for the content of statistics and data science collaborations. The $Q_1Q_2Q_3$ approach explicitly emphasizes the importance of the qualitative context of a project, as well as the qualitative interpretation of quantitative findings. We provided guidance for implementing each stage of the approach and presented data evaluating the effectiveness of teaching the $Q_1Q_2Q_3$ approach to beginning collaborators. We recommend that the $Q_1Q_2Q_3$ method be taught to both current students of statistics and professional statisticians. We believe that the $Q_1Q_2Q_3$ approach provides statisticians with an approach for successfully contributing to research, policy, and business decisions to transform evidence into action for the benefit of society.

# 7. Acknowledgements

This material is based upon work supported by the National Science Foundation under Grant No. 1955109 and Grant No. 2022138 for the projects, "IGE: Transforming the Education and Training of Interdisciplinary Data Scientists (TETRDIS)" and "NRT-HDR: Integrated Data Science (Int dS): Teams for Advancing Bioscience Discovery." This work was also partially supported by the United States Agency for International Development under Cooperative Agreement #7200AA18CA00022. We would like to thank the Laboratory for Interdisciplinary Statistical Analysis (LISA) student collaborators and domain experts, as well as students from the Statistical Collaboration course at the University of Colorado Boulder.

Olubusoye, O. E., Akintande, O. J., & Vance, E. A. (2021). Transforming Evidence to Action: The Case of Election Participation in Nigeria. *CHANCE*, *34*(3), 13–23. https://doi.org/10.1080/09332480.2021.1979807

Paul, R., & Elder, L. (2013). *Critical Thinking: Tools for Taking Charge of Your Professional and Personal Life* (2nd edition). Ft Pr.

Pearl, J. (1995). Causal Diagrams for Empirical Research. *Biometrika*, *82*(4), 669–688. https://doi.org/10.2307/2337329

Petocz, P., & Reid, A. (2010). On Becoming a Statistician—A Qualitative View. *International Statistical Review*, *78*(2), 271–286. https://doi.org/10.1111/j.1751-5823.2010.00101.x

Vance and Smith (2021)*.* Asking Great Questions: Part of a Theory of Communication in Interdisciplinary Collaborations*, JSM Section on Statistical Consulting,* https://par.nsf.gov/servlets/purl/10310043

Vance, E. A. (2020). Goals for Statistics and Data Science Collaborations. *JSM Proceedings, Statistical Consulting Section*. https://par.nsf.gov/biblio/10227760-goals-statistics-data-science-collaborations

Vance, E. A., Alzen, J. L., & Smith, H. S*.* (2022a). Creating Shared Understanding in Statistics and Data Science Collaboration*s*. *Journal of Statistics and Data Science Education.* https://www.tandfonline.com/doi/pdf/10.1080/26939169.2022.2035286

Vance, E. A., & Love, K. (2021). Building Statistics and Data Science Capacity for Development. *CHANCE*. https://doi.org/10.1080/09332480.2021.1979810

Vance, E. A., & Smith, H. S. (2019). The ASCCR Frame for Learning Essential Collaboration Skills. *Journal of Statistics Education*, *27*(3), 265–274. https://doi.org/10.1080/10691898.2019.1687370

Vance, E. A., Trumble, I.M., Alzen, J.L., Smith, H.S. (2022b). Asking Great Questions. *In press.*

Wiens, A., Lovejoy, H. B., Mullen, Z., & Vance, E. A. (2022). A modelling strategy to estimate conditional probabilities of African origins: The collapse of the Oyo Empire and the transatlantic slave trade, 1817–1836. *Journal of the Royal Statistical Society: Series A (Statistics in Society)*, 1-24. https://doi.org/10.1111/rssa.12833

21

Yu, B., & Kumbier, K. (2020). Veridical data science. *Proceedings of the National Academy of Sciences*, *117*(8), 3920–3929. https://doi.org/10.1073/pnas.1901326117